\documentclass[12pt,preprint]{aastex}
\usepackage{epsf}

\def\etal{{\it et al.}}
\def\eg{{\it e.g.,~}}
\def\cf{{\it cf.,~}}
\def\ie{{\it i.e.,~}}
\def\Htwo{${\rm H_2}~$}

\def\H{{\rm H}}
\def\He{{\rm He}}
\def\K{~{\rm K}}
\def\kms{~{\rm km~s^{-1}}}
\def\cm3{~{\rm cm^{-3}}}
\def\yrs{~{\rm yrs}}

\def\Msun{~{\rm M}_{\sun}}
\def\Zsun{~{\rm Z}_{\sun}}
\def\uR{ u_{\rm R}}
\def\lsim{\mathrel{  
        \raise0.3ex\hbox{$<$}\kern-0.75em{\lower0.65ex\hbox{$\sim$}}}}
\def\gsim{\mathrel{
        \raise0.3ex\hbox{$>$}\kern-0.75em{\lower0.65ex\hbox{$\sim$}}}}

\slugcomment{draft of \today}

\shorttitle{THERMALLY UNSTABLE PRIMORDIAL CLOUDS}
\shortauthors{BAEK, KANG, AND RYU}

\begin{document}

\title{Two-Dimensional Axisymmetric Collapse of \\
         Thermally Unstable Primordial Clouds}

\author{Chang Hyun Baek, Hyesung Kang}
\affil{Department of Earth Sciences, Pusan National University, Pusan
            609-735, Korea}
\email{chbaek@comet.es.pusan.ac.kr, kang@uju.es.pusan.ac.kr}

\and

\author{Dongsu Ryu}
\affil{Department of Astronomy \& Space Science, Chungnam National
             University, Daejeon 305-764, Korea}
\email{ryu@canopus.chungnam.ac.kr}

\begin{abstract}
We have performed two-dimensional hydrodynamic simulations of the
collapse of isolated axisymmetric clouds condensing via radiative cooling
in a primordial background gas. In order to study the development of
the so-called ``shape-instability'', we have considered two types of
axisymmetric clouds, oblate and prolate clouds of various sizes and
with axial ratios of $0.5 \leq {R_{\rm c,R}} /{R_{\rm c,z}} \leq 2$.
We find that the degree of oblateness or prolateness is enhanced
during the initial cooling phase. But it can be reversed later,
if the initial contrast in cooling times between the cloud
gas and the background gas is much greater than one.
In such cases an oblate cloud collapses to a structure composed
of an outer thin disk and a central prolate component.
A prolate cloud, on the other hand, becomes a thin cigar-shape structure
with a central dense oblate component. The reversal of shape
in the central part of the cooled clouds is
due to supersonic motions either along the disk plane in the case of
oblate clouds or along the symmetry axis in the case of prolate clouds.
For a background gas of $T_h=1.7\times 10^6$K and $n_h=0.1 \cm3$ in
a protogalactic halo environment, the mean density of the cloud gas
that has cooled to $10^4$K increases to $100 n_h$ or so, in our
simulations where nonequilibrium cooling is adopted and the background
gas cools too. The spherical Jeans mass of such gas is estimated
to be about $M_J \sim 5\times10^{7}\Msun$.
In order for cloud mass to exceed the Jeans mass and at the same
time in order for the thermal instability to operate, the initial cloud
size should be around $1 - 1.5 l_{\rm cool}$ where
$l_{\rm cool}$ is the cooling length.
\end{abstract}


\section{Introduction}

In many areas of astrophysics, the {\it thermal instability} is often
invoked to explain the condensation of cold dense clouds 
out of a hot background medium \citep[\eg][]
{fie65, gold70, def70, sch72, fall85, bal89, vaz00, koy02, kri02}.
In the simplistic picture of the thermal instability,
the overdense region surrounded by the hotter background
undergoes a ``quasi-static compression'' in near pressure equilibrium.
This ``near-equilibrium'' case of the evolution, however,
is valid only when the cloud is small enough to adjust to pressure change
faster than it cools -- its size must be much smaller than the distance 
that sound wave travels in a cooling time, $l_{\rm cool}$.
The collapse of thermally unstable clouds in either X-ray cluster
cooling flows or protogalactic halos has been studied previously
by numerical simulations
\citep[\eg][and references therein]{dav88, bri90, kang00}. These
simulations showed that a spherically symmetric cloud cools and undergoes
a supersonic compression, if the cloud size  is $R_{\rm c} \sim l_{\rm cool}$.
Such supersonic compression leads to a central density increase two
to three orders of magnitude higher than what is expected
from the isobaric compression.
On the other hand, a small cloud of $R_{\rm c} < l_{\rm cool}$
cools isobarically and undergoes a quasi-static compression,
while a big cloud of $R_{\rm c} > l_{\rm cool}$ cools nearly isochorically
with only small density increase.

While most previous studies on the thermal instability
considered the collapse of one-dimensional (1D)
spherically symmetric clouds, \citet{bri90}
simulated the collapse of a non-spherical, elongated blob in the
two-dimensional (2D) polar geometry and showed the development
of a ``shape instability''.
As the perturbation is compressed by the background pressure, 
the compression wave travels the same distance with the sound speed
(\ie $l \sim c_{\rm s} \cdot t$) along both the major and minor axes,
and the induced infall velocity field is not radial. 
This causes the compressed region 
to be more elongated and enhances the degree of non-sphericity. 
They showed that an oblate cloud with an initial axial ratio
of $b/a = 0.447$ collapses to a flat pancake, and argued that
the evolution of oblate clouds becomes similar to that of 
1D plane-parallel collapses.
From this simulation one can deduce that the shape instability would also
occur in a prolate cloud, resulting in a thin rod-shape condensation.
In \citet{kang00}, we studied the effects of different geometries
by 1D plane-parallel and 1D spherical symmetric simulations of
thermally unstable clouds using a PPM (Piecewise Parabolic Method) 
hydrodynamic code with self-gravity and radiative cooling.
We found that isotropic compression leads to much higher central density,
accompanied by accelerated radiative cooling, in the spherically
symmetric case, while the density increases only to the isobaric ratio of
$\rho_c/\rho_h = (T_h \mu_c /T_c \mu_h)$ in the plane-parallel case.
Here $\mu$ is the mean molecular weight, and the subscripts
$c$ and $h$ stand for cool and hot, respectively.
Thus, we expect that the cloud shape is an important factor
in the collapse and evolution of thermally unstable clouds. 
In this contribution, we explore in detail how {\it non-spherical}
perturbations in a {\it primordial} environment evolve under
the influence of the thermal instability.

In many models of globular cluster (GC) formation, 
dense protoglobular cluster clouds (PGCCs)
are supposed to exist in pressure equilibrium with the hot gas
in protogalaxies \citep[\eg][]{gunn80, bro91, kum93}.
The model by \citet{fall85}, which has been most widely adopted
to explain the existence of such PGCCs, relies on the thermal instability
for the formation of PGCCs in a protogalactic halo. 
In our study the parameters have been adjusted so the simulations for
the nonlinear development of the thermal instability are applicable
to the condensation of PGCCs, although the generic results
should hold to the thermal instability of any objects.
We have ignored self-gravity, because the gravitational time scale is
much longer than the cooling time scale during the early collapse stage
of PGCCs. But the self-gravity should be important in the gravitational
fragmentation of PGCCs during the later evolutionary stage when
the clouds have cooled down to $10^4$K.

In next section, we describe our model and numerical method.
The simulation results are presented in \S III, followed by the
conclusions in \S IV.

\section{HYDRODYNAMIC SIMULATIONS}

\subsection{Isolated Clouds in an Uniform Background Halo}

We expect that inside a
protogalactic halo, density perturbations on a wide range of length scales
exist and flow motions are likely turbulent. While the thermal instability
under such realistic global pictures can be explored later
\citep[\cf][]{vaz00, kri02}, here we first take a much simpler and local
approach. We consider isolated overdense clouds embedded in a hot,
uniform, and static background gas of $T_h = 1.7\times 10^6$K and
$n_h = 0.1 \cm3$. This temperature corresponds to that of an isothermal
sphere with circular velocity $V_c=220\kms$, representing the halo
of disk galaxies like the Milky Way Galaxy. $n_h$ is the background density
of hydrogen nuclei. The value of $n_h = 0.1 \cm3$ is chosen as a fiducial
value, because then spheres of radius
$R_c \sim l_{\rm cool}$ would have mass scales relevant for GC formation
(see \S 2.3). We assume the ratio of He/H number densities is 1/10, so that
the gas mass density is given by $\rho_h = (2.34\times 10^{-24}{\rm g})n_h$. 

The initial density of the overdense clouds, $n_{\rm cloud}$, is assumed
to decrease gradually from the center to edge,
\begin{equation}
n_{\rm cloud} = n_h \left[1 + \delta \cos\ \biggl\{ {\pi\over2}
\left[({R \over R_{\rm c,R}})^2 +({ z \over R_{\rm c,z}})^2 \right]^{1/2}
\biggr\} \right] 
\end{equation}
for $(R/R_{\rm c,R})^2+(z/R_{\rm c,z})^2 \le 1$,
where $R_{\rm c,R}$ and $R_{\rm c,z}$ are the cloud radii along the $R$ 
and $z$ axes, respectively. The ratio of $R_{\rm c,R} / R_{\rm c,z}$
determines the shape of the initial clouds (see \S 2.5).
The initial temperature throughout the clouds is set by the 
isobaric condition, \ie $T_{\rm cloud}=T_h (n_h/n_{\rm cloud}$). 
The amplitude of initial density perturbations, $\delta$, is
a free parameter that determines the density contrast between
the cloud center and the background. 
If this amplitude is linear (\ie $\delta \ll 1$) and if there are no
heat sources that balance the radiative cooling of the background gas,
then both the cloud and the background gas cool together and 
the thermal instability would not have enough time to grow. 
Because the cooling time scales as 
$t_{\rm cool} \propto (1+\delta)^{-2}$,
the contrast in $t_{\rm cool}$ between the cloud and 
the background is small for small $\delta$.
In real protogalaxies, however, the halo gas would be heated 
by possible energy sources such as supernova explosions, shocks, and etc. 
If the background gas maintains a high temperature owing to
those energy inputs, perturbations would grow approximately 
under the isobaric condition until they become nonlinear.
Also these heating processes likely induce turbulent flow motions
and non-linear density fluctuations in the halo medium as well.
In numerical simulations the overall evolution proceeds faster for 
larger values of $\delta$ during the linear phase, but it becomes almost
independent of the initial values of $\delta$, once perturbations
become non-linear. 
Since we do not include any background heating processes 
in our simulations, we need to have a reasonably large density contrast
in order to see the nonlinear growth and to expedite the simulations. 
Thus, we begin with $\delta=1$ for all models, which in fact
would be consistent with turbulent nature of the halo medium. 

\subsection{Radiative Cooling Rates}

The key idea of the GC formation model based on the thermal instability,
originally suggested by \citet{fall85}, is that the characteristic mass
scale of GCs, $M_c \sim 10^6 \Msun$, can be explained by the imprinting
of the Jeans mass of the gas clouds at $T=10^4$K that have cooled from
a hot halo gas in pressure equilibrium.
If the clouds were allowed to cool well below $10^4$ K in a time scale
shorter than the free-fall time scale, they would not retain the memory
of the imprinted mass scale.
So in this work we consider the cases where the radiative cooling is 
ineffective below $10^4$K.
If the halo gas had been enriched by the metals from the first objects,
or if \Htwo molecules have formed efficiently via gas phase reactions,
then the gas would have cooled well below $10^4$ K before 
the Jeans mass was imprinted \citep{shkang87}. 
Thus, we consider a primordial gas with $\H$ and $\He$ only,
and we assume that the formation of \Htwo molecules is
prohibited due to UV radiation from central AGNs or diffuse
background radiation \citep{kang90}, resulting in zero cooling
for $T<10^4$K.

Here, we define the cooling rate as
\begin{equation}
\Lambda(T,n_H) = L(T) n_H^2,
\end{equation}
where $\Lambda$ is the energy loss rate per unit volume 
and $n_H$ is the number density of hydrogen nuclei.
In general, the cooling rate coefficient, $L(T)$, is a function of
temperature as well as ionization fractions that can be dependent on
the thermal and ionization history of gas.
When a hot gas cools from $T>10^6K$, in particular,
it recombines out of ionization equilibrium,
because the cooling time scale is shorter than the recombination
time scale \citep{shkang87}.
So in order to calculate the non-equilibrium cooling rate accurately,
the time-dependent equations for ionization fractions should be solved,
which can be computationally expensive.
Fortunately, however, the non-equilibrium cooling rate for the gas 
{\it cooling under the isobaric condition} becomes a function of 
temperature only, if the initial temperature is high enough 
to ensure the initial ionization equilibrium (\eg $T>10^6$K)
and if only two-body collisional processes are included.  
In that case, $L(T)$ can be represented
by a tabulated form as a function of temperature only.
We adopt, as our standard cooling model, the non-equilibrium radiative 
cooling rate for a zero-metalicity, optically thin gas 
that is calculated by following the non-equilibrium collisional
ionization of the gas cooling from $10^{7.5}$K to $10^4$ K under 
the isobaric condition \citep{suth93}.
We set $L(T)=0$ for $T<10^4$K and, in addition, set the minimum
temperature at $T_{\rm min}=10^4$K.

In order to explore the effects of different cooling rates, we have
also calculated several models with the following cooling rates,
in addition to the standard cooling model (NEQm0): 
1) the CIEm0 model: the collisional ionization
equilibrium cooling for a zero-metalicity gas, 
2) the NEQm1 model: the non-equilibrium cooling rate for a gas with
the metalicity,  $Z=0.1\Zsun$.
The NEQm1 model is not consistent with our assumption of zero cooling
rate below $T=10^4$, but has been included for comparison.
Figure 1 shows the cooling rate coefficients for these cooling models.

As the central part of the cloud cools,
a steep temperature gradient develops between the cloud and
hot background medium and the thermal conduction can become operative there. 
\citet{bri90} showed, however, that their simulation results 
differ by $\le$3\% when the reduced thermal conductivity \citep{gry80}
was included. 
Thus we ignore the thermal conduction in our calculations, although
possibly it may be important.

\subsection{Cooling Time and Length}

We define the cooling time as
\begin{equation}
t_{\rm cool} = {{\epsilon} \over {\Lambda}} = {{3/2 n_{\rm tot} k T}
\over {\Lambda}},
\end{equation}
where $\epsilon = p /(\gamma -1)$ is the internal energy per unit volume,
and $n_{\rm tot}$ is the number density of ions and electrons. 
With the standard NEQm0 cooling model, 
the cooling time for the background halo gas is
\begin{equation}
t_{\rm cl,h}= 2.0 \times 10^{7} \yrs \cdot {( {n_h \over {0.1 \cm3}})^{-1}}
\end{equation}
with $T_h = 1.7 \times 10^6$K. Figure 1 shows the cooling time for
a gas cooling under the isobaric condition (\ie $n_{\rm tot} T$= constant)
for three cooling models. The cooling time for the gas at the cloud center
is given by
\begin{equation}
t_{\rm cl,c} = {t_{\rm cl,h}\over(1+\delta)^2} = {t_{\rm cl,h}\over4}
\end{equation}
with $n_{\rm cloud}=n_h (1+\delta)$, $T_{\rm cloud}=T_h /(1+\delta)$
and $\delta=1$. 
We note the cloud center cools in a time scale of $\sim 0.5 t_{\rm cl,h}$,
while the background halo cools in $ \sim 2 t_{\rm cl,h}$.

We define the cooling length as the distance over which the sound wave
of the hot halo gas travels within one cooling time of the perturbed gas,
\begin{equation}
l_{\rm cool} =c_h \cdot t_{\rm cl,c} = 1.05 {\rm kpc} \cdot
{({ n_h \over{0.1  \cm3}})^{-1}},
\end{equation}
where $c_h=198\kms$ is the sound speed of the hot gas of
$T_h = 1.7 \times 10^6$K.
The dynamics of radiatively cooling clouds are characterized by the
cloud size relative to the cooling length \citep{fall85, dav88, kang00}.

The mass contained within a cloud of radius, $R_c=l_{\rm cool}$, is
\begin{equation}
M_{\rm cool} = {4\pi \over 3} l_{\rm cool}^3 \rho_h
= 1.75\times 10^7 \Msun \cdot  {({ n_h \over{0.1  \cm3}})^{-2}}.
\end{equation}
For $n_h=0.1 \cm3$, the clouds with the size $R_c \sim (0.4-1) l_{\rm cool}$
have the mass scales of $M_{\rm cloud} \approx 10^6 - 10^7 \Msun$
that are relevant for PGCCs. 
Because of the $n_h^{-2}$ dependence,
for the background halo density that is much larger or much smaller than
our fiducial value, the characteristic cooling mass would be too small
or too big, respectively, for the cooled perturbations to become PGCCs.

\subsection{Numerical Method}

The gasdynamical equations for an axisymmetric system in the cylindrical
coordinate system, $(R,z,\phi)$, including radiative cooling are written as
\begin{equation}
{\partial \rho \over \partial t} + {1 \over R} {\partial \over {\partial R}}
(R\rho \uR) + {\partial \over \partial z} ( \rho u_z)= 0,
\end{equation}
\begin{equation}
{\partial (\rho \uR) \over  \partial t} + {1 \over R}{\partial \over
\partial R}(R \rho \uR^2) + {\partial \over \partial z}(\rho \uR u_z) -
{\partial P \over \partial R} = 0,
\end{equation}
\begin{equation}
{\partial (\rho u_z) \over  \partial t} + {1 \over R}{\partial \over
\partial R}(R \rho \uR u_z) + {\partial \over \partial z}(\rho u_z^2) -
{\partial P \over \partial z} = 0,
\end{equation}
\begin{equation}
{\partial (\rho e) \over \partial t} + {1 \over R} {\partial \over
{\partial R}} [R \uR (\rho e + P)] + {\partial \over \partial z}
[u_z (\rho e + P)] = -\Lambda,
\end{equation}
where $e=\epsilon/\rho + (1/2)u^2$ is the total energy of the gas
per unit mass, $R=\sqrt{x^2+y^2}$, $u=\uR^2+u_z^2$,
and the rest of the variables have their usual meanings.

To solve the hydrodynamic part, we have used an Eulerian,
grid-based hydrodynamics code based on the
``Total Variation Diminishing (TVD)'' scheme \citep{ryu93}.
The cylindrical geometry version has been used for 2D simulations.
For 1D comparison simulations, the spherical geometry version has been used.
The TVD scheme solves a hyperbolic system of gasdynamical conservation
equations with a second-order accuracy. Multidimensionality is
handled by the Strang-type dimensional splitting \citep{strang68}.

After completion of the hydrodynamic part updating hydrodynamical
quantities from the time step $t^n$ to $t^{n+1}$, radiative cooling
is applied to the thermal energy as a separate part.
If we were to update the thermal energy density by the following
explicit scheme,
\begin{equation}
\epsilon^{n+1} = \epsilon^n - \Lambda \Delta t
= \epsilon^n ( 1 - {\Delta t \over t_{\rm cool}^{n+1/2}}),
\end{equation}
the time step size should be smaller than the cooling time,
that is, $\Delta t < t_{\rm cool}^{n+1/2}$. Here ${t_{\rm cool}^{n+1/2}}$
is estimated from the time averaged hydrodynamic quantities,
$(q^n + q^{n+1})/2$.
This explicit integration scheme would be extremely expensive,
when the cooling time scale is much shorter than the hydrodynamical
time scale. For that reason, we rewrite the cooling part
of the thermal energy equation as
\begin{equation}
{{d \ln(\epsilon)} \over dt} = - { \Lambda \over \epsilon }
= - { 1 \over t_{\rm cool}}.
\end{equation}
In our numerical code 
we integrate this equation as
\begin{equation}
\epsilon^{n+1} = \epsilon^n \cdot \exp(- {{\Delta t_h}
\over {t_{\rm cool}^{n+1/2}}}),
\end{equation}
assuming that $t_{\rm cool}$ is constant over the hydrodynamic time scale
$\Delta t_h$ \citep[\eg][]{LeV97}.
Strictly speaking, Equation (14) converges to Equation (12) only when
$\Delta t_h$ is much smaller than $t_{\rm cool}$. 
When $\Delta t_h\gsim t^{n+1/2} _{\rm cool}$, however,
the gas quickly looses most of the thermal energy via radiation
and the temperature approaches to the specified minimum value ($T_{\rm min}$)
during one hydrodynamic time step, 
if we were to integrate Equation (12) with a small timestep size of 
$t_{\rm cool}$ {\it for many steps}. 
But this behavior is emulated reasonably well by integrating Equation (14) 
with $\Delta t_h$ {\it for one step}, which effectively lowers the gas 
temperature to $T_{\rm min}$. 
So the end results of both integration schemes would be qualitively
similar, although could be quantitatively somewhat different.
Especially at different spatial grid resolutions, the detail thermal
history of the gas could be different during the cooling phase, but once
cooled the final structure should be roughly similar. 
With this integration scheme, gas cooling can be followed,
although approximately, with the hydrodynamic time steps, even when
the cooling time scale is shorter than the hydrodynamic time scale.

Our simulations start at $t=0$ with the clouds at rest in pressure
equilibrium ($P_{\rm halo} = P_{\rm cloud}$), and cease at $t=2t_{\rm cool}$
when the {\it background} gas has cooled to $10^4$K.
The standard mirror condition has been used for the reflecting boundaries
at $R=0$ and $z=0$, while it has been assumed that flows are continuous
across the outer boundaries. We have used only one quadrant
of the cylindrical coordinate system.

In our simulations the clouds cool and collapse due to the compression
by the background pressure to the size that is $1/10$ or so of the initial
cloud radii.
In order to study the detail structure of the collapsed clouds,
we have devised a special grid that consists of an {\it inner fine zone}
with uniform grid spacing and an {\it outer coarse zone} with expanding
grid spacing. The inner zone has $1000^2$ uniform cells with
$ \Delta R = \Delta z = L_{\rm fine}/1000$, where $L_{\rm fine}=
1.6 l_{\rm cool}$. 
The outer zone has $1240^2 - 1000^2$ cells covering the rest of
the simulated region outside the inner fine zone, with the cell
spacing that increases outwards as
$\Delta R^i /\Delta R^{i-1} = \Delta z^i / \Delta z^{i-1} = 1.05$ for
$i =1001,...1240$.
With the expanding grid the outer boundaries are located far away from
the central cloud where most activity occurs.

The physical variables are expressed in units of the following
normalization both in the numerical code
and in the plots presented below:
$t_0 =t_{\rm cl,h} = 2.0\times 10^7$ years; $r_0 = l_{\rm cool} = 1.05$kpc; 
$u_0 = l_{\rm cool} / t_{\rm cl,h} = 50.8 {\rm kms}^{-1}$;
$\rho_0 = 2.34 \times 10^{-24}{\rm g} \cdot n_h$; $P_0= \rho_0 u_0 ^2=
6.03\times 10^{-12}{\rm erg~cm^{-3}} $.

\subsection{Shape Parameter}

In order to quantify how the cloud shape evolves,
we define the ``shape parameter'' as
\begin{equation}
S = {R_{\rm e,R} \over R_{\rm e,z}},
\end{equation}
where $R_{\rm e,R}$ and $R_{\rm e,z}$ are the ``effective'' radii
along the $R$ and $z$-axes, respectively.
Here, the ``effective'' radius is defined as the radius
where the gas density decreases to a half of the central density,
that is, $\rho(R_e) = (1/2)\rho_c$.
The initial values of the shape parameter for models considered are
$S_{\rm i} = 1$ for spherical clouds, $S_{\rm i} = 5/6$ and 1/2
for prolate clouds, and $S_{\rm i} = 6/5$ and 2 for oblate clouds.
The second column of Table 1 shows the initial values of the shape
parameter for each model.

\section{SIMULATION RESULTS}

\subsection{Spherical Symmetric Calculations}

We have first calculated spherical symmetric collapses both with 1D and
2D codes, in order to show their behavior and to compare it with that
of non-spherical collapses, as well as to test the performance of the 2D code.
According to our previous 1D simulations which employed a different code
(the PPM code) \citep{kang00}, 
the evolution of spherical clouds collapsing via the thermal instability
can be classified by the cloud size as follows:
1) $(R_c/l_{\rm cool}) < 0.2$, the {\rm isobaric} compression regime,
2) $0.2\lsim (R_c/l_{\rm cool}) \lsim 1-1.5$, the {\it supersonic}
compression regime, and 3) $(R_c/l_{\rm cool}) \gsim 1-1.5$, the
{\it isochoric} cooling regime.

The following four different sizes have been considered in our
spherical symmetric calculations:
three {\it supersonic} cloud models, 
S11 with $R_{\rm c} = 0.4l_{\rm cool}$,
M11 with $R_{\rm c} = 0.8l_{\rm cool}$,
and L11 with $R_{\rm c} = 1.6l_{\rm cool}$,
one {\it isochoric} cloud model, 
X11 with $R_{\rm c} = 3.2l_{\rm cool}$ (see Table 1).
The largest cloud model, X11, has a cloud mass too large for PGCCs,
but it has been included for comparison. 
We first note that the results from the current 1D simulations 
are consistent with those from our previous simulations in \citet{kang00}, 
although different numerical codes are used. 
While the CIEm0 cooling model was adopted in \citet{kang00},
here we have adopted the standard NEQm0 cooling model 
which has lower peaks around both the H and He Ly $\alpha$ line emissions 
(see Figure 1). As a result, gas cools less rapidly and
central density increases to lower values in the current simulations.

The results from 2D simulations of spherical collapses are mostly
similar to those from 1D simulations.
Figure 2 shows the distributions of the gas density,
the velocities in the $R$ and $z$ directions and the pressure
along the diagonal line of $R = z$ from the 2D simulation for
the L11 model with $R_{\rm c} = 1.6l_{\rm cool}$.
The distributions are sampled at $t/t_{\rm cl,h}=0.2$ (solid line),
$0.6$ (dotted line), $1.0$ (dashed line), $1.4$
(long dashed line) and $1.8$ (dot-dashed line).
As the central gas cools and loses pressure,
the high pressure background compresses the cloud 
and induces an infall flow for $t/t_{\rm cl,h} \lsim 0.6$.
The flow velocity increases up to $ (u/c_h) \sim 0.36$,
which is much larger than the sound speed of the cooled gas,
$(c_c /c_h) \sim 0.08$. So the accretion flow is supersonic.
After $t=0.6t_{\rm cl,h}$, the infall flow bounces back
as an accretion shock and the shock moves out.
Then, the cloud expands outward slowly due to 
the high pressure at the central region. 
One can see that the central density increases up
to $\rho_c/\rho_o \sim 10^6$ just before the shock bounces back,
and then decreases to $\rho_c/\rho_o\sim 10^{3.5}$
as the collapsed cloud expands.
The shape of the collapsed clouds from 2D simulations becomes
slightly rectangular, especially for the models with small sizes
(see Figure 8 below). The deviation from the spherical symmetry,
measured with the shape parameter $(S-1)$, ranges from less
than 1\% (in the X11 model) to $\sim20$\% (in the S11 model).

\subsection{2D Axisymmetric Calculations}

For the non-spherical, axisymmetric clouds,
four different sizes are considered, as for the spherical clouds: 
$R_{\rm c,max} = 0.4 l_{\rm cool}$ for small clouds,  
$R_{\rm c,max}=0.8l_{\rm cool}$ for medium size clouds, 
$R_{\rm c,max} = 1.6 l_{\rm cool}$ for large clouds, and
$R_{\rm c,max} = 3.2 l_{\rm cool}$ for very large clouds.
Here, $R_{\rm c,max} = \max[R_{\rm c,R},R_{\rm c,z}]$. The initial
parameters of the model clouds considered are summarized in Table 1.

Figures 3 and 4 show the time evolution of the density distribution
of the large prolate cloud, the L56 model with $S_{\rm i}=5/6$,
and the large oblate cloud, the L65 model with $S_{\rm i}=6/5$, respectively.
Figure 5 shows the flow velocity field superimposed on the density
contour maps of the L56 (left panels) and L65 (right panels) models
at $t=0.8 t_{\rm cl,h}$ (top panels) and
at $t=1.6 t_{\rm cl,h}$ (bottom panels).
As gas cools catastrophically, the clouds implode near the center 
and supersonic anisotropic accretion flow is induced.
The compression continues until $t \approx 0.8t_{\rm cl,h}$ when
infall flow is reflected at the center and an accretion shock forms.
This ``shock formation time'' depends on the initial cloud size,
as the infall flow turns around later in larger clouds.
At the shock formation time, the infall flow field is strongly
anisotropic, preferentially parallel to the $z=0$ plane for the prolate
cloud and parallel to the $z$-axis for the oblate cloud.
As a result, the degree of prolateness or oblateness is enhanced.
So the prolate cloud collapses to a very thin rod shape,
while the oblate cloud collapses to a flat disk, at
$t \approx 0.8t_{\rm cl,h}$ in the L56 and L65 models. 
The central density peaks at the time
of shock formation, and then slowly decreases afterwards.
After that time, the accretion shock halts the infall flow, as shown
in the bottom panels of Figure 5. Inside the accretion shock,
there exist residual infall motions with the preferred direction
different from that of early accretion before $t=0.8 t_{\rm cl,h}$,
that is, along the $z$-axis for the prolate model and along the $z=0$
plane for the oblate model. 
So now the contraction proceeds along the $z$-axis toward the center, 
which in turn generates the radial outflow along the $z=0$ plane 
inside the cooled cloud in the prolate model.
In the oblate model, on the other hand,
the contraction occurs along the $z=0$ plane toward the center
and induces the bipolar outflow along the z-axis. 
This ``secondary'' contraction leads to a central bulge
that has a shape reverse to the initial shape of the cloud.
Thus, the prolate cloud results in a structure
that consists of a cigar-shaped outer component
and a central {\it oblate} component. 
On the other hand, the oblate cloud collapsed to a structure 
of a pancake-shaped outer component and a central {\it prolate} component.

In the 2D simulation of \citet{bri90}, they considered an oblate cloud
($S_{\rm i}\approx 2.2$) with small density contrast, $\delta=0.1$, 
in a background with $T_h=10^7$K and $n_h=0.1{\rm cm}^{-3}$, hotter
than ours.  They showed that the cloud collapses to a flat structure
with density enhancement of $\sim 100$ at $t\sim 3$ in units of their
cooling time. Because their simulation started with a small $\delta$,
it ended just after the formation of a nonlinear pancake-like structure
that corresponds to the structure at $t\sim 0.9 t_{\rm cl,h}$ in our
simulation (see Figure 4).  As a result, they did not see the development
of the inner central prolate condensation. Whether collapses end at
the cigar/pancake formation stage or they continue to form the inner
oblate/prolate components depends mainly on the cloud size ratio,
$R_c/l_{\rm cool}$, and the contrast of cooling times,
$t_{\rm cl,c}/t_{\rm cl,h}$. Thus, these two parameters along
the initial shape parameter, $S_{\rm i}$, determine the dynamics and
mass distribution of the collapsed clouds. 

Figures 6 and 7 show the evolutionary sequences for the medium size 
prolate cloud, the M12 model with $S_i=1/2$, and the medium size oblate 
cloud, the M21 model with $S_i=2$, respectively.
Due to a higher degree of initial non-sphericity, these models
display much stronger shape instability than the L56 and L65 models.
The shock formation time is $t \approx 0.6t_{\rm cl,h}$, slightly earlier
than that for the L56 and L65 models because of smaller sizes.
The inner components due to the secondary contraction grow less 
significantly, compared to those of the L56 and L65 models.

To illustrate how the initial shape parameter and cloud size
affect the evolution, we plot in Figure 8 the density contour
maps at $t =1.6t_{\rm cl,h}$ for the models among listed
in Table 1 which adopt the standard cooling.
For all models, the regions bounded by $[-0.5 R_{\rm c,R},
0.5 R_{\rm c,R}] \times [-0.5 R_{\rm c,z},0.5 R_{\rm c,z}]$
are shown. So for example, [-0.4, 0.4] $\times$ [-0.4, 0.4] in units
of $l_{\rm cool}$ is shown for the M models.
Except in the very large clouds (the X models), the shape reversal
is observed in the central part of cooled clouds; that is, the 
formation of the inner oblate (prolate) component in the prolate
(oblate) models. Spherical models are shown to demonstrate how well
the spherical symmetry is conserved in 2D simulations (\S 3.1). 
It is interesting to see that the collapses of the
clouds with initially even 20\% deviation from the spherical
symmetry (\ie M56, M65, L56 and L65) are very different from the 
spherical collapses. For a given value of $S_i$, the models with
smaller cloud sizes have more significant inner
components, since the shock formation occurs earlier.
For a given cloud size, on the other hand, the models with $S_i$'s
closer to one form more significant inner components.
In the very large cloud models, the sound crossing time along the long
axis ($R_{\rm c,max}=3.2l_{\rm cool}$) is much longer than the cooling
time. Hence, the spherical cloud (the X11 model) cools and collapses 
without forming an accretion shock. 
However, the X12 and X21 models have the minor-axis radius, 
$R_{\rm c,min} = 1.6 l_{\rm cool}$, so they are able to 
collapse supersonically along the minor-axis forming an accretion 
shock at $t \approx t_{\rm cl,h}$. As a result,
by $t = 1.6t_{\rm cl,h}$ the prolate cloud (the X12 model)
has been developed into a rod with the central density
$\rho_c/\rho_h \sim 10^{2.7}$, while the oblate cloud (the X21 model)
into a pancake with $\rho_c/\rho_h \sim 10^{2.3}$.
As in the simulation by \citet{bri90}, the inner central components
do not display the shape reversal by the time $t \approx 1.6t_{\rm cl,h}$
in these models. If we adopt a higher initial density contrast or keep
the background gas at constant pressure, however, even the X models
could have developed the inner central components with the reversed shape. 

In order to look at the density enhancement in a quantitative way, we
present in Figure 9 the density line profiles of the collapsed clouds
at $t/t_{\rm cl,h}=1.8$ for the initially prolate (left panels) and
oblate (right panels) clouds.
The dotted (dashed) lines show the profiles along the
$R$-axis ($z$-axis), while the solid lines show the profiles of the 2D
spherical models with the same size (\ie the S11-X11 models) along
the diagonal line of $R=z$ as a function of the radial distance
$r=\sqrt{R^2+z^2}$. In the prolate models the dashed lines reveal the thin
rod-shape component along the $z$-axis, while the dotted lines show
the inner oblate component along the $R$-axis, except in the X12 model
where the inner oblate component does form. 
In the oblate models, on the other hand, the dotted lines show the flat
pancake-shape component along the $R$-axis, while the dashed lines reveal
the inner prolate component along the $z$-axis, except in the X21 model.
Once again the shape reversal is observed in the central region
($r\la 0.1-0.2 l_{\rm cool}$) of the cooled clouds, except in the X
models. The outer rod-shape component in the prolate models and
the outer pancake-shape component in the oblate models have
the density close to the isobaric ratio,
$\rho_{\rm cool} / \rho_{\rm h} \approx ( {T_h \mu_c}/ {T_c\mu_h } ) 
\approx 10^{2.5}$.
The inner oblate/prolate components have the mean density 
somewhat higher ($\rho_{\rm cool} /\rho_h \approx 10^{2.5-3.5}$) 
in the M and L models.
As mentioned before, the central density peaks at the time of shock
formation and then decreases afterwards as the clouds expand, so
the density distribution changes in time.
If we compare the central density of different models at a given time
after the inner shape reversal appears, the oblate models have
higher values than the prolate models, except in the X models. 

Figure 10 shows the evolution of the shape parameter, $S$,
for the models with $S_i=1/2$ (prolate) and 2 (oblate). 
During the first ``shape instability'' stage, the shape parameter is
determined by the cigar-shape component or the pancake-shape component.
During the ``secondary infall'' stage after the shock formation, however,
it represents the shape of the inner components in most models (except
in the S12 model). In the oblate clouds (right panels),  
as the clouds collapse to flat pancakes, the value of $S$
increases to $\gg 1$ until the shock formation time. 
Such time when $S$ reaches the maximum values scales as the sound
crossing time ($t_{\rm sc}=R_{\rm c,min}/c_h$), which is proportional to
the cloud size.
Afterwards, the value of $S$ decreases, and it becomes smaller than one 
due to the formation of the inner {\it prolate} components, except
in the model X21. In the prolate clouds (left panels),
the value of $S$ decreases to $\ll 1$ as the clouds collapse to thin
rod shapes. After the shock formation the value of $S$ increases,
and it becomes greater than one in the M12 and L12 models as the inner
{\it oblate} components grow.
In the small S12 model, the inner oblate component and the outer
cigar-shaped component have the similar density, so the effective radius
along the $z$-axis represents the length of the cigar component
rather than the radius of the inner oblate component,
resulting in $S=R_{\rm e,R}/R_{\rm e,z} < 1$. For this model, we have
also calculated the second shape parameter $S_{100}$, which is defined
as the ratio of the radii where $\rho(R_{100})= 100 \rho_h$.
As shown in Figure 10, this second shape parameter becomes $S_{100}>1$
at $t\ge 1.6 t_{\rm cl,h}$, indicating that the inner oblate component
indeed forms in the S12 model. In the X models, $S$ remains either
less than one or greater than one, as expected.

\subsection{Cloud Mass and Mean Density}

In the Fall and Rees model for the formation of PGCCs where the
thermal instability is assumed to proceed quasi-statically,
the ``critical mass'' defined for an isothermal sphere confined 
by an external pressure \citep{mcc57}
\begin{equation}
M_{cr} =1.18\left({k T_c \over {\mu m_H }}\right)^2 G^{-3/2}p_h^{-1/2}
\end{equation}
was adopted as the minimum mass for gravitationally unstable clouds, 
For $T_c=10^4 \K$ and $p_h=3.28\times 10^{-11} {\rm dyne~cm^{-2}}$,
$M_{cr} = 2.8\times 10^6 \Msun$.
However, this simple picture should be modified for the following reasons:
1) the collapse is not quasi-static and the infall flows can become
supersonic, so the compressed clouds are bound by the ram pressure of
the infall flows rather than the background pressure, 2) the cooled
compressed clouds (PGCCs) may have turbulent velocity fields (see Figure 5),
3) the mass distribution of PGCCs can be very complex, rather than
spherical (see Figures 3-8).

As an effort to obtain a better estimation on the PGCC mass, we have
estimated the Jeans mass of the cooled gas in our simulations as follows.
First, we have calculated the mass of the gas with $T=10^4$K,
$M_4$. The right panels of Figure 11 show how $M_4$ increases
with time in our models. Most of the cloud gas cools to $10^4$K
by $t=1.4t_{\rm cl,h}$, so by this time $M_4$ becomes comparable to
the initial cloud mass, $M_{\rm cloud}$.
We present the values of $M_4$ at $t=1.4t_{\rm cl,h}$ in the fifth
column of Table 1. Next, from the mass and volume of the cells with
$T = 10^4$K, we have calculated the mean density of the central region,
$<\rho_4>$. The values of $<\rho_4>$ are shown in the left panels of
Figure 11. Although the central density can increase to values much
higher than the isobaric compression ratio of $10^{2.5}$, the mean
density, $\rho_4$, increases only up to that ratio and then decreases,
as the background gas cools and the cooled cloud expands. The values
of $<\rho_4>$ at $t=1.4t_{\rm cl,h}$ are listed in the last column
of Table 1.

Finally, although it may not be a good approximation either because of
the reasons listed above, we have computed the Jeans mass of
an isothermal uniform sphere with temperature $T_c$ and density
$\rho_c$ as \citep{spitz79}
\begin{equation}
M_J=\rho \lambda_J^3 =5.46 \left({k T_c \over {\mu m_H G}}\right)^{3/2}
{\rho_{c}}^{\, -1/2}
\end{equation}
by adopting $T_c = 10^4$K and $\rho_c =<\rho_4>$. The values of
such Jean masses are plotted in the right panels of Figure 11, 
and listed in the 8th column in Table 1.
In the spherical model, the cooled gas in the L11 cloud has mass greater
than the Jeans mass, \ie $M_4 \gsim M_J$. In the non-spherical models,
the total mass of the cooled gas in the L56, L65, and L21 clouds exceeds
the Jeans mass. These clouds might become gravitationally unstable
after cooling to $10^4$K. In those models, the clouds might imprint
the Jeans Mass of $M_J \sim 5 \times 10^7 \Msun$. Assuming a star
formation efficiency of order 10 \%, this mass estimate is a bit too
large for the characteristic mass scale of the GC mass distribution.
This mass scale, however, is much larger than the previously estimation
by \citet{kang00}. It is because the cloud density increases only to
$\rho_c \approx 100 \rho_h$ in the current simulations where the slower
nonequilibrium cooling rate has been adopted. In any case, our estimation
for the gravitationally unstable mass scale should be taken to be very
rough, and would depend on the model parameters including the
density and temperature of the background.
In addition, self-gravity has been ignored in our simulations, which
is expected to be important in the later stage of the cloud evolution,
as pointed in \S 1. Because the X11, X12 and X21 models
either have small density enhancement or have extremely non-spherical
mass distribution, we argue that it would not be useful to make
the comparison between $M_4$ and $M_J$ for those models.   

\subsection{Different Cooling Model}

Finally, we have calculated the L56 and L65 models with different
coolings, to see their effects. They are labeled as EL56 and EL65
for the models with the CIEm0 cooling and as ML56 and ML65
for the models with the NEQm1 cooling (see also Table 1).
Figure 12 shows the density contour maps at the shock formation time
($t=0.8 t_{\rm cl,h}$) and at a later time ($t=1.4 t_{\rm cl,h}$).
In the ML56 and ML65 models where cooling is enhanced by about a factor
of 3.5 at $T\sim10^6$K  due to metals, the cloud gas has cooled to
$T=10^4$ K already at $t=0.2t_{\rm cl,h}$ and the compression wave
steepens into a pair of shocks (reverse and forward) by $t=0.8t_{\rm cl,h}$.
This produces an elongated ring-like structure in the 2D image, which is
in fact a dense elongated shell in 3D. This structure along with the
pair of shocks collapses at the center and then an accretion shock
bounces back before $t=t_{\rm cl,h}$. The gas of the EL56 and EL65 models
with the CIEm0 cooling cools faster, and so we see the resulting structure
is more compact than that in the models with the standard NEQm0 cooling.
As a result, the density enhancement reaches $<\rho_4> = 360-860 \rho_h$
by $t=1.4 t_{\rm cl,h}$, which are greater than those found in the models
with the standard cooling. So the spherical Jeans mass of the cooled
clouds becomes $\sim 10^7 \Msun$, which is more consistent with
the characteristic mass scale of GCs (see Table 1). 

\section{SUMMARY}

In many astrophysical problems, a two-phase medium may develop by
the formation of cold dense clouds via the thermal instability.   
In order to explore how the non-spherical shape affects the collapse 
of thermally unstable clouds, we have performed 2D 
hydrodynamical simulations in the cylindrical geometry
with the radiative cooling rate for a primordial gas.
Although we have selected for our simulations
a physical environment  relevant to a protogalactic halo,
the overall simulation results can be applied to the collapse of
thermally unstable clouds of any physical scales, 
as long as the cooling curve has the pertinent characteristics 
for the thermal instability.

The collapse of non-spherical clouds depends mainly on the following
factors.\hfil\break
1) The ratio of the cloud size to the cooling length, 
$R_c/l_{\rm cool}$: As shown in the previous 1D spherical simulations
\citep[\eg][]{dav88, bri90, kang00}, clouds of $R_c \sim l_{\rm cool}$
undergo a supersonic compression, resulting in high density enhancements.
Small clouds with $R_c < l_{\rm cool}$ cool nearly isobarically through 
a quasi-static compression, while large clouds with $R_c > l_{\rm cool}$
cool nearly isochorically. Hence, we have focused on the clouds with
$R_c \sim l_{\rm cool}$.\hfil\break
2) The initial density contrast between the cloud and the background,
$(1+\delta)$: Note that the contrast in cooling times,
$t_{\rm cl,c}/t_{\rm cl,h}$, is initially proportional to $(1+\delta)^{-2}$ for
an isobaric perturbation. Since we have not included any heating source
that maintains the temperature  of the background gas, the growth of
nonlinear structures depends on the value of $\delta$.  For example,
with $\delta \ll 1$, only the first stage of nonlinear growth can develop
before the background gas itself cools down. With $\delta \sim 1$,
however, more complex structures form after the initial emergence of
the shape instabilities. So we have started our simulations with
$\delta = 1$.\hfil\break
3) The degree of non-sphericity, $S=R_{\rm c,R}/R_{\rm c,z}$:
It is obvious that the degree of deviation from the spherical
symmetry determines the dynamics of the infall flows and the mass
distribution of the collapsed clouds. So we have considered both
the prolate clouds with $S_{\rm i}=1/2$ and 5/6 and the oblate clouds
with $S_{\rm i}=6/5$ and 2, in addition to the spherical clouds
($S_{\rm i}=1$).

The collapse in our simulations can be described by two distinct
stages: the first {\it shape instability stage} during which
the non-sphericity grows due to the initial infall, and the
{\it secondary contraction stage} during which the infall occurs
predominantly along the direction perpendicular to the initial infall
flows and a ``shape reversal'' occurs.
Even with initially only 20 \% deviation from the spherical
shape (\ie $S_{\rm i}=5/6$ or 6/5), a strong shape instability occurs,
so the prolate clouds are compressed to thin rods and the oblate clouds
are compressed to flat pancakes due to strongly anisotropic infall flows.
The degree of prolateness or oblateness, however, is enhanced only
during the initial shape instability phase up to the formation of
accretion shocks. 
Afterwards, secondary infall motions are induced,
dominantly along the $z$-axis for the prolate clouds and along
the $z=0$ plane for the oblate clouds.
This secondary contraction parallel or perpendicular to the $z$-axis
induces, within the cooled clouds, 
the radial outflows in the prolate models or the bipolar
outflows in the oblate models, resulting in
the inner central bulges with the mass distribution opposite
to the initial shape. 
As a result,
initially prolate clouds collapse to a system that consists
of an outer cigar-shaped component and a central oblate component,
while initially oblate clouds collapse to a system that consists
of an outer pancake-shaped component and a central prolate component.

The central density of the collapsed clouds increases until accretion
shocks form at the end of the first shape instability stage.
And then it gradually decreases as the clouds expand, since the 
central pressure is higher than the background pressure.
The central density in our simulations has
turned out to be much lower than that in the previous
simulations. In the secondary contraction stage,
the mean density of the outer pancake or thin-rod components is
similar to the background density times the isobaric compression
ratio of $(T_h\mu_c/T_c\mu_h)=10^{2.5}$, while that of the inner
components reaches only up to an order of magnitude higher than that.
We note that the density enhancement depends on the radiative cooling
rate. It would be higher in the simulations with larger cooling rates.
We have adopted as the standard cooling model, the cooling rate of
a primordial gas based on the non-equilibrium ionization fraction
tabulated by \citet{suth93}. This cooling model has lower rates near
H and He Ly $\alpha$ line emission peaks than the cooling model
based on the collisional ionization equilibrium which was adopted in
our previous 1D simulations \citep{kang00}. So we have found
smaller density enhancements in the current simulations.

For the protogalactic halo environment considered here,
$\rho_h = 2.34 \times 10^{-25} {\rm g} \cdot \cm3$ and
$T_h = 1.7 \times 10^6$K, the spherical Jeans mass of the cooled
clouds is about $M_J \sim 5\times10^7 \Msun$. This mass scale is
somewhat large for the mass of PGCCs which fragment to form GCs.
But this is based on a very rough estimation which would depend
on model parameters. In addition, in a realistic halo environment,
the halo gas may be heated by the stellar winds, supernova
explosions, shock waves and etc. If the halo can maintain
the high temperature and continue to compressed the PGCCs,
the density of PGCCs would have increased more, resulting in
a smaller Jeans mass. Finally, we have considered here a static halo
of uniform density. However, protogalactic halos are likely
clumpy and turbulent, and the PGCCs may have formed in such
environment. We leave all these issues, along with extension
into 3D, to future works.

\acknowledgments{
This work was supported by grant No. R01-1999-00023
from the Korea Science \& Engineering Foundation.
The numerical calculations were performed through
``The 2nd Supercomputing Application Support Program'' of KISTI
(Korea Institute of Science and Technology Information).
We thank the anonymous referee for clarifying comments.}

\clearpage

\begin{table*}
\begin{center}
{\bf Table 1.}~~Initial Parameters for Model Clouds\\
\vskip 0.3cm
\begin{tabular}{ lrrrrrrrrr }
\hline\hline
Model & $S_i$ & $R_{\rm c,R}$/$l_{\rm cool}$ & $R_{\rm c,z}$/$l_{\rm cool}$ & $R_{\rm c,max}$ & $M_{\rm cloud}$ & $M_{4}$~\tablenotemark{a} & $M_{J}$~\tablenotemark{a} & $<\rho_{4}>$~\tablenotemark{a}\\
~ & ~ & ~ & ~ & (kpc) & ($10^6\Msun$) &  ($10^6\Msun$) &  ($10^6\Msun$) & ($2.34\times 10^{-24}{\rm g} \cdot n_h$)\\

\hline
S11 &1.0 &0.4 &0.4  &0.421 & 1.489 & 1.156 & 45.222 & 96.125 \\
S12 &0.5 &0.2&0.4  &0.421 & 0.376 & 0.455 & 47.744 & 86.236\\
S21 &2.0 &0.4 &0.2 &0.421 & 0.745 & 0.856 & 49.797 & 79.271\\
\hline
M11 &1.0 &0.8 &0.8  &0.842 & 11.858 & 8.940 & 32.831 & 182.367\\
M12 &0.5 &0.4 &0.8  &0.842 & 2.979 & 2.687 & 47.934 & 85.557\\
M56 &5/6 &0.67 &0.8 &0.842 & 8.243 & 8.059 & 65.272 & 46.139\\
M65 &6/5 &0.8 &0.67 &0.842 & 9.881 & 7.935 & 65.407 & 45.949\\
M21 &2.0 &0.8 &0.4  &0.842 & 5.929 & 5.928 & 50.757 & 76.304\\
\hline
L11 &1.0 &1.6 &1.6  &1.68 & 94.622 & 71.112 & 58.688  & 57.074\\
L12 &0.5 &0.8 &1.6  &1.68 & 23.715 & 19.752 & 40.654   &  118.941 \\
L56 &5/6 &1.33 &1.6 &1.68 & 56.741 & 55.689 & 39.360   &  126.890 \\
L65 &6/5 &1.6 &1.33 &1.68 & 78.852 & 61.853 & 45.436   &   95.218 \\
L21 &2.0 &1.6 &0.8  &1.68 & 47.310 & 40.701 & 35.067   &  159.857 \\
EL56 &5/6 &1.33 &1.6 &1.68 & 56.741 & 64.934 & 6.653   &  855.259 \\
EL65 &6/5 &1.6 &1.33 &1.68 & 78.852 & 65.678 & 10.307  &  356.416 \\
ML56 &5/6 &1.33 &1.6 &1.68 & 56.741 & 92.528 & 46.025  &   93.201 \\
ML65 &6/5 &1.6 &1.33 &1.68 & 78.852 & 105.559& 35.390  &  157.635 \\
\hline
X11 &1.0 &3.2 &3.2 &3.36 & 756.981 & 571.840 & 193.961 & 5.225\\
X12 &0.5 &1.6 &3.2  &3.36 & 189.721 & 157.458 & 47.869 & 85.789 \\
X21 &2.0 &3.2 &1.6  &3.36 & 378.481 & 286.662 & 65.393 & 45.969 \\
\hline
\tablenotetext{a}{Calculated at $t=1.4t_{\rm cl,h}$.}
\end{tabular}

\end{center}
\end{table*}

\clearpage

\begin{figure}
\plotone{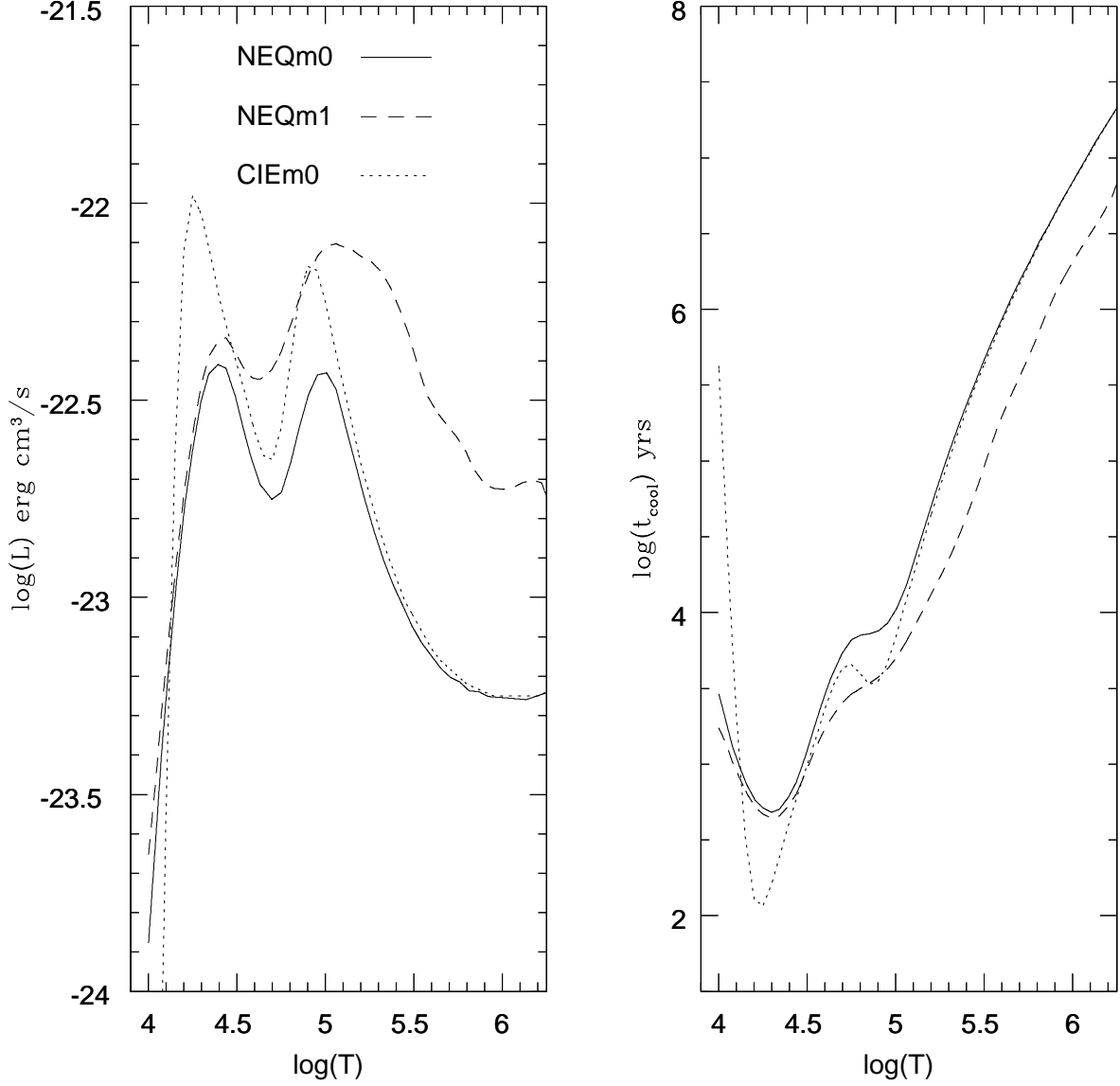}
\caption{
{\it Left panel}: Cooling rate coefficient, $L(T)= \Lambda/n_H^2$, 
for the standard cooling model (NEQm0, solid line), the collisional
ionization equilibrium cooling model for a zero-metalicity
gas (CIEm0, dotted line), and the non-equilibrium cooling model for
a gas with the metalicity, $Z=Z_\sun/10$ (NEQm1, dashed line) (see text).
{\it Right panel}: Cooling time,
$t_{\rm cool} = {\epsilon}/ {\Lambda}$, 
for a gas cooling from $T_h=1.7\times 10^6$K with $n_h=0.1 {\rm cm^{-1}}$
initially under the isobaric condition. The same line type is
used for different cooling models as in the cooling rate coefficient.}
\end{figure}

\begin{figure}
\plotone{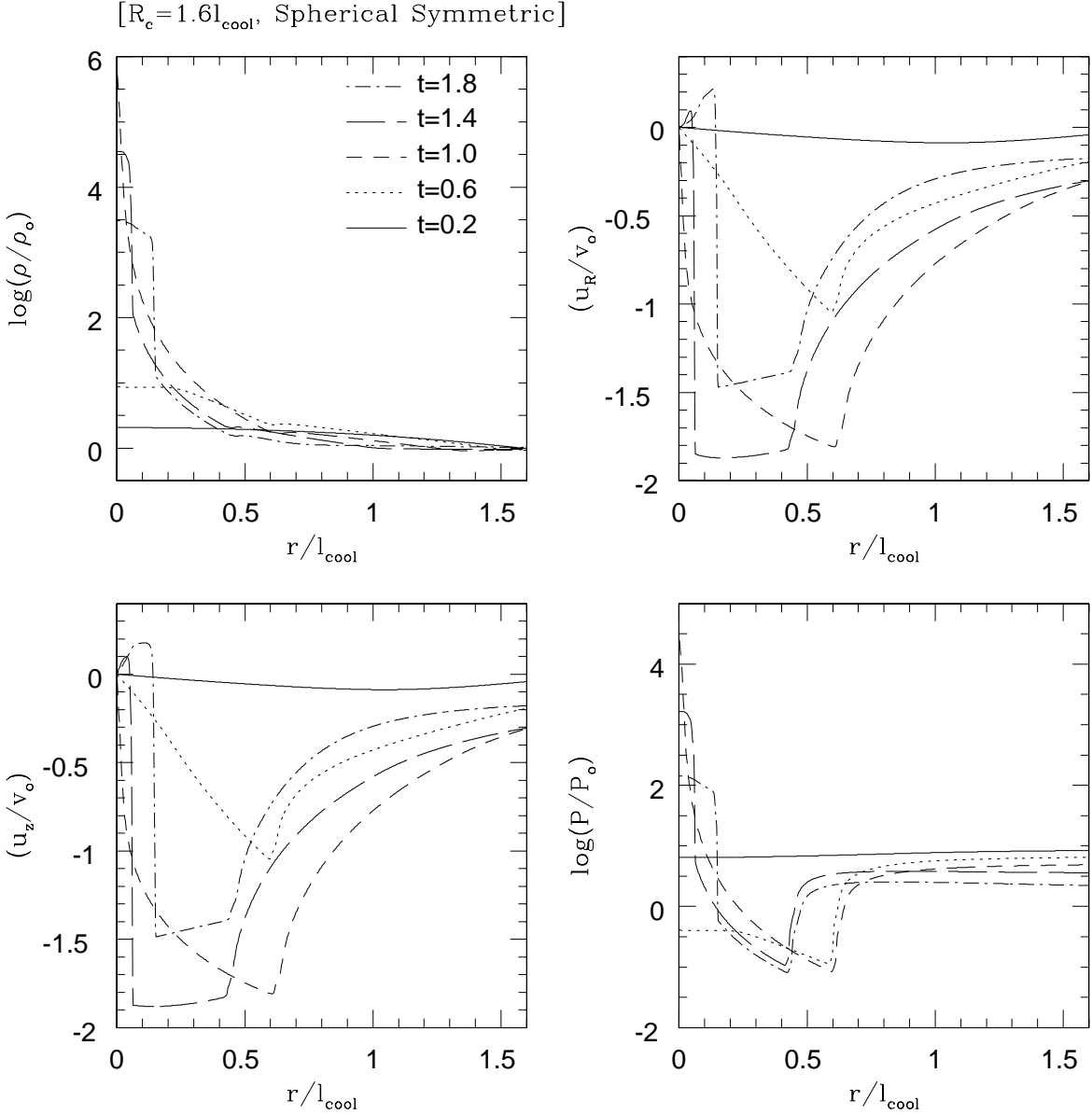}
\caption{
Time evolution of the gas density, $\rho$, the velocities,
$\uR$, $u_z$, and the pressure, $P$, at $t/t_{\rm cl,h}=$
0.2 (solid), 0.6 (dotted), 1.0 (dashed), 1.4 (long dashed)
and 1.8 (dot and dashed line) for the spherical cloud
with $R_{\rm c}=1.6 l_{\rm cool}$ (the L11 model).
Distributions along the diagonal line of $R=z$ in 2D box are shown,
as a function of the radial distance $r= \sqrt{R^2+z^2}$.}
\end{figure}
\begin{figure}
\plotone{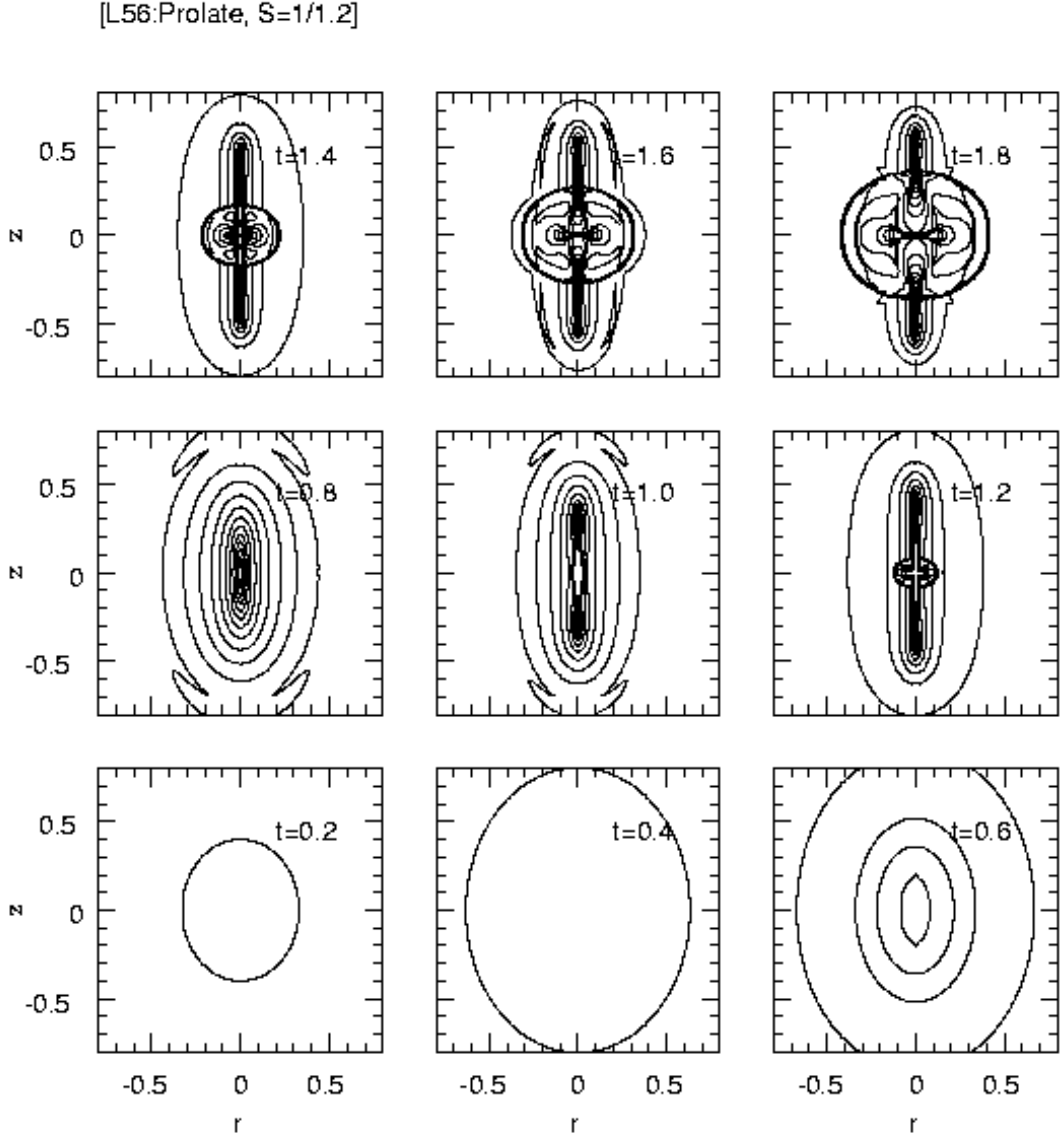}
\caption{
Time evolution of the density distribution in the prolate cloud
with $S=5/6$ and $R_{\rm c,max} = 1.6 l_{\rm cool}$ (the model L56).
The contour levels for $(\rho / \rho_0) = 2^i$ ($i=0, 1, \cdots 10$) are shown.
Full images are generated from the simulation data covering one quadrant
of the whole space by using the reflection symmetry along the $z=0$
plane and along the $z$-axis.}
\end{figure}
\begin{figure}
\plotone{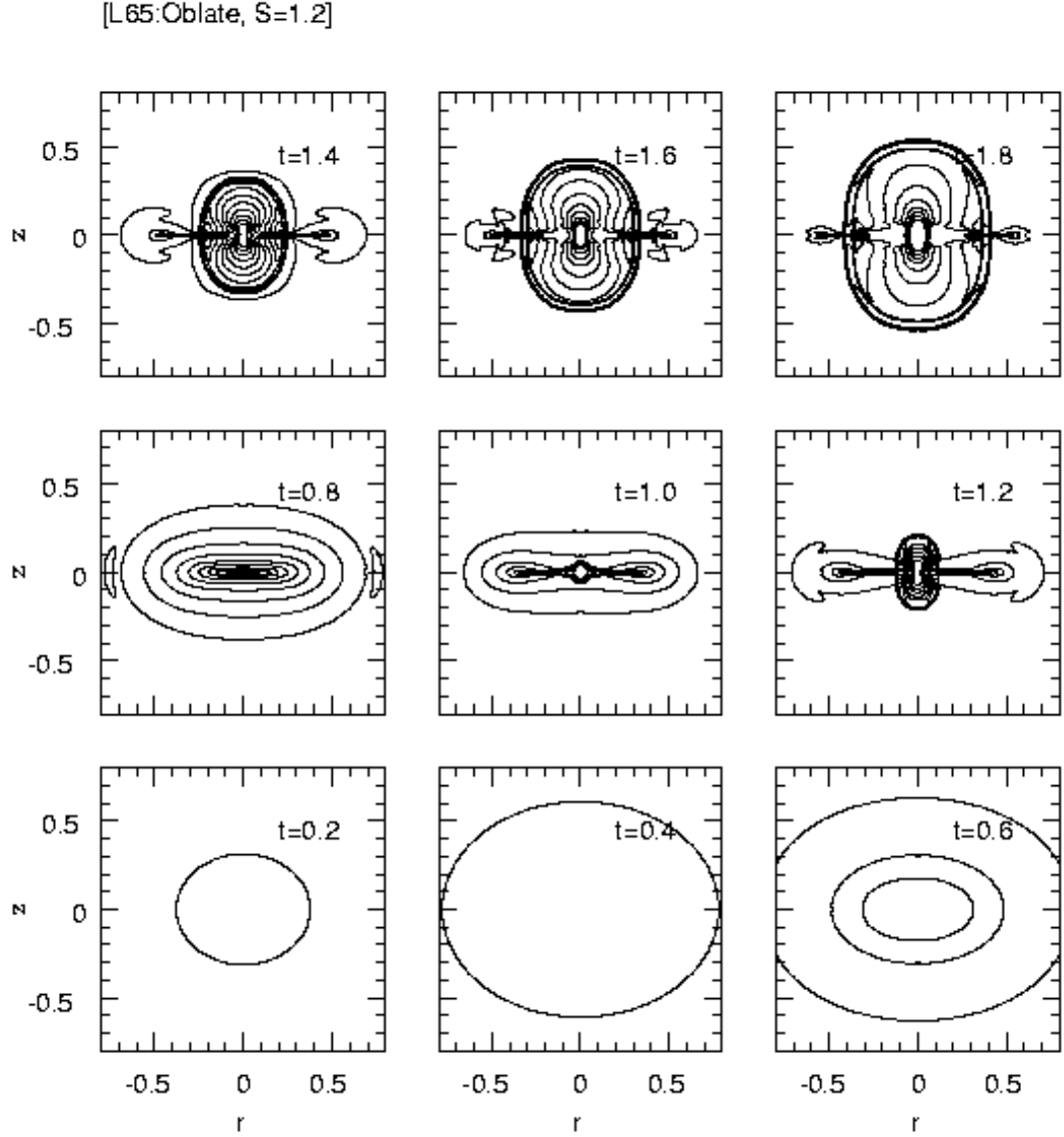}
\caption{
Same as in Figure 3 except for the oblate cloud with $S=6/5$ and
$R_{\rm c,max} = 1.6l_{\rm cool}$ (the model L65).}
\end{figure}
\begin{figure}
\plotone{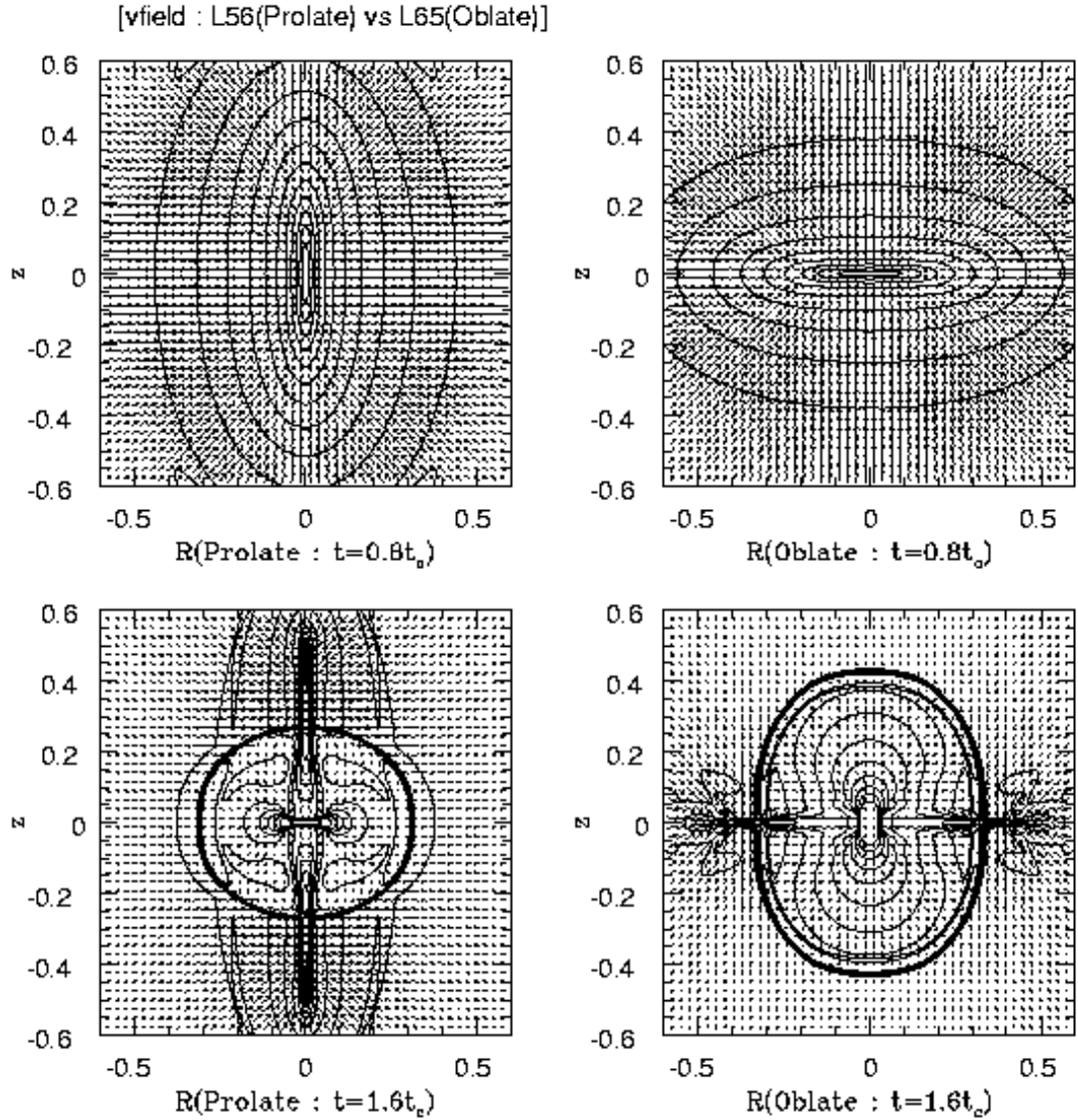}
\caption{
The velocity field overlapped on the density contour
map for the prolate model L56 (left panels) and
for the oblate model L65 (right panels)
at $t=0.8 t_{\rm cl,h}$ (top panels) and
at $t=1.6 t_{\rm cl,h}$ (bottom panels).}
\end{figure}
\begin{figure}
\plotone{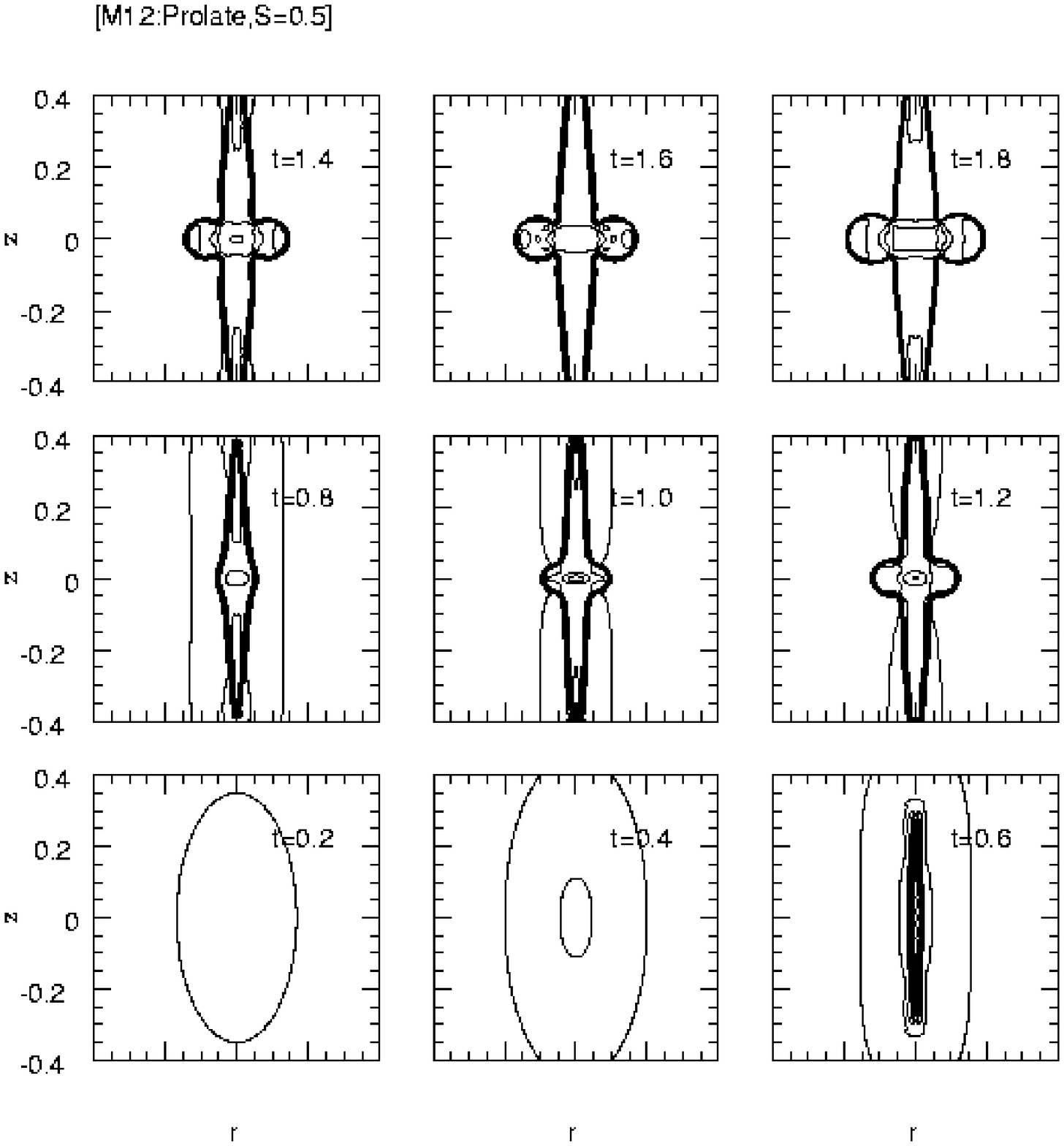}
\caption{
Same as in Figure 3 except for the prolate cloud with $S=1/2$ and
$R_{\rm c,max} = 0.8l_{\rm cool}$ (the model M12).}
\end{figure}
\begin{figure}
\plotone{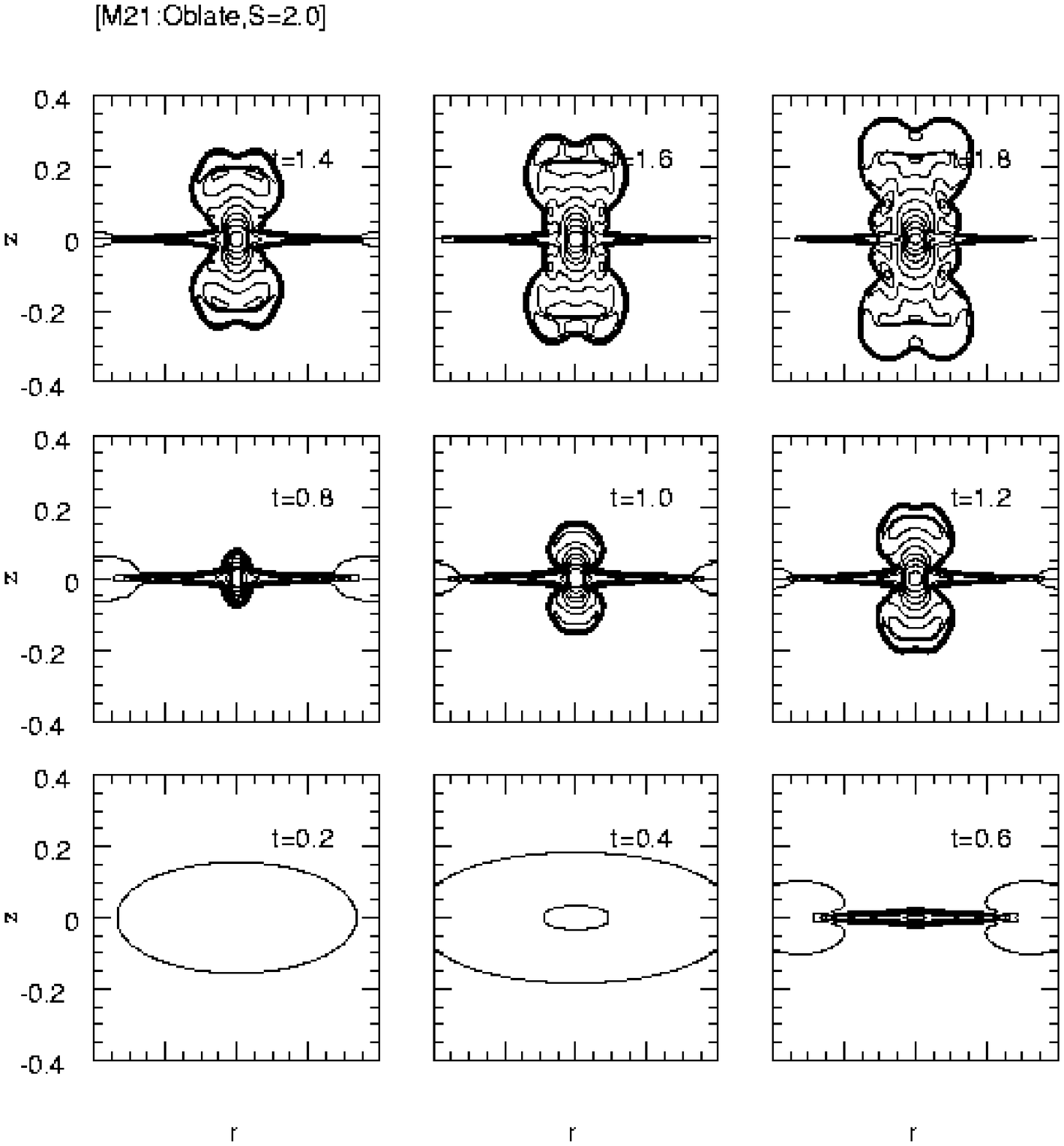}
\caption{
Same as in Figure 3 except for the oblate cloud with $S=2/1$ and
$R_{\rm c,max} = 0.8l_{\rm cool}$ (the model M21).}
\end{figure}
\begin{figure}
\plotone{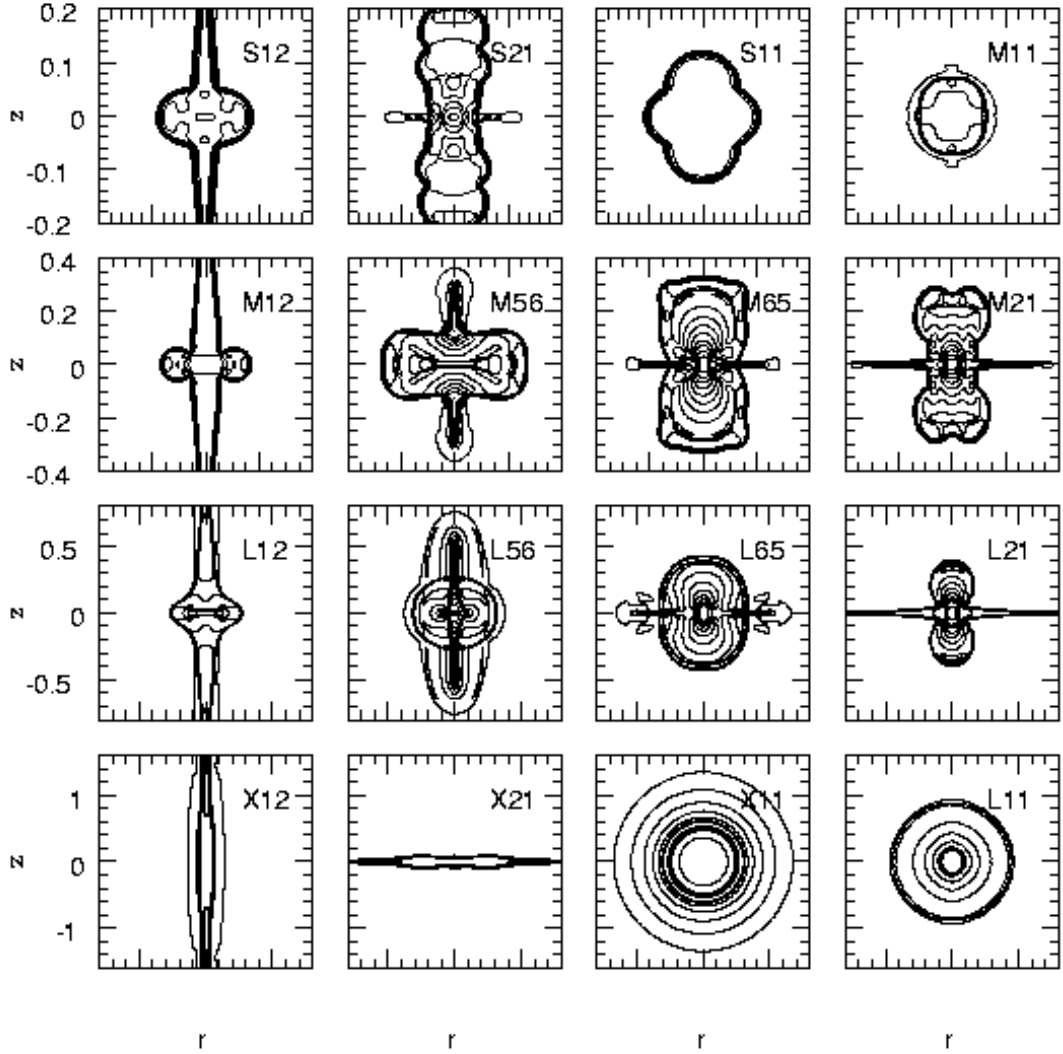}
\caption{
Density contour maps at $t = 1.6 t_{\rm cl,h}$ for all the models
with the standard cooling. The contour levels for
$(\rho / \rho_0) = 2^i$ ($i=0, 1, \cdots 10$) are shown.
Each map is labeled with the model name (see Table 1).}
\end{figure}
\begin{figure}
\plotone{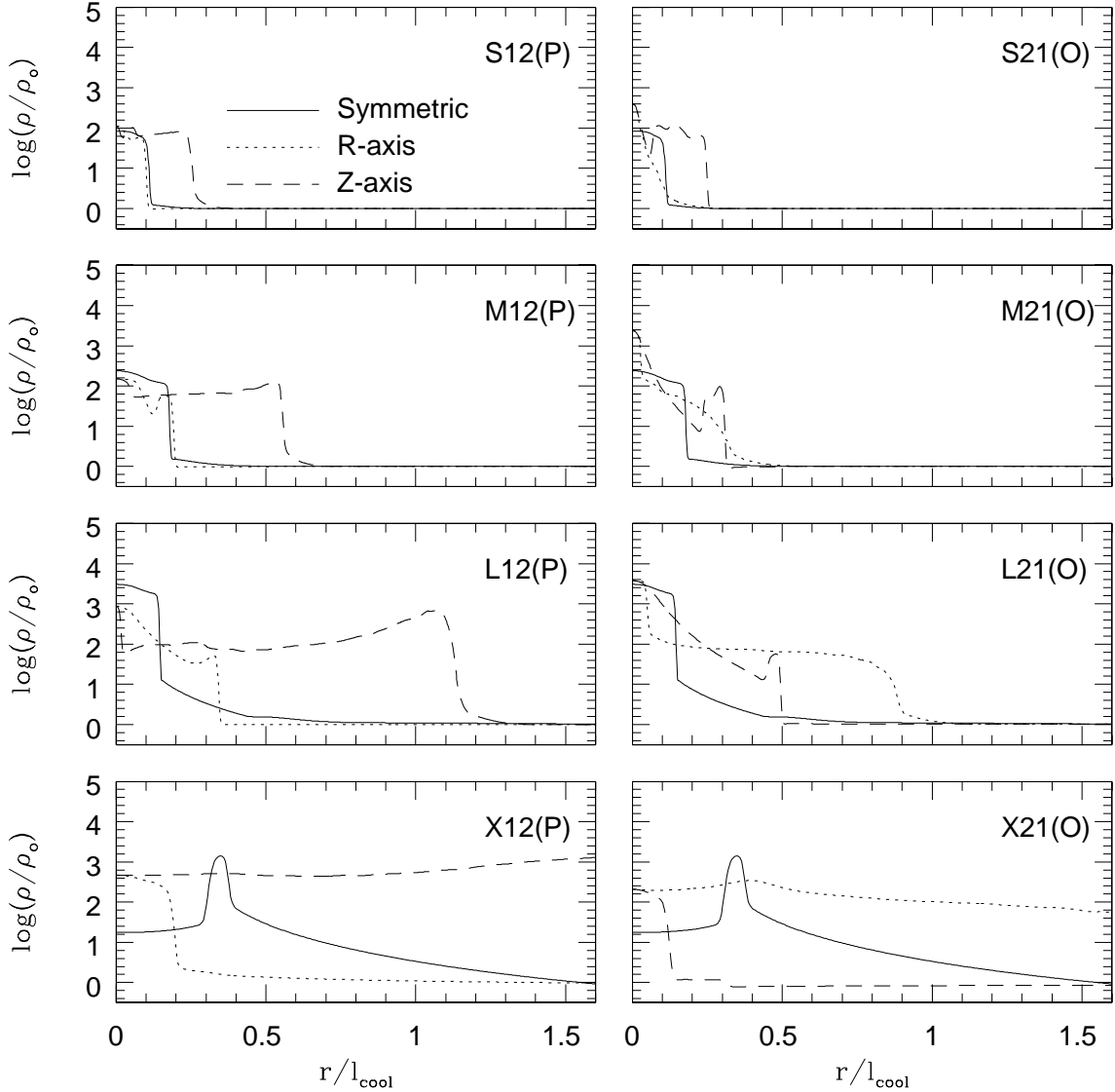}
\caption{
Density line profiles of the cooled clouds for the S12 (top-left),
S21 (top-right), M12 (upper middle-left) and M21 (upper middle-right)
models at $t=1.2t_{\rm cl,h}$, and for the L12 (lower middle-left), L21
(lower middle-right), X12 (bottom-left), and X21 (bottom-right) models
at $t=1.4t_{\rm cl,h}$. The dotted and dashed lines show the profiles
along the $R$ and $z$-axes, respectively. The results of 1D
spherical simulations (the S11, M11, L11, and X11 models) are shown
with the solid lines for comparison.}
\end{figure}
\begin{figure}
\plotone{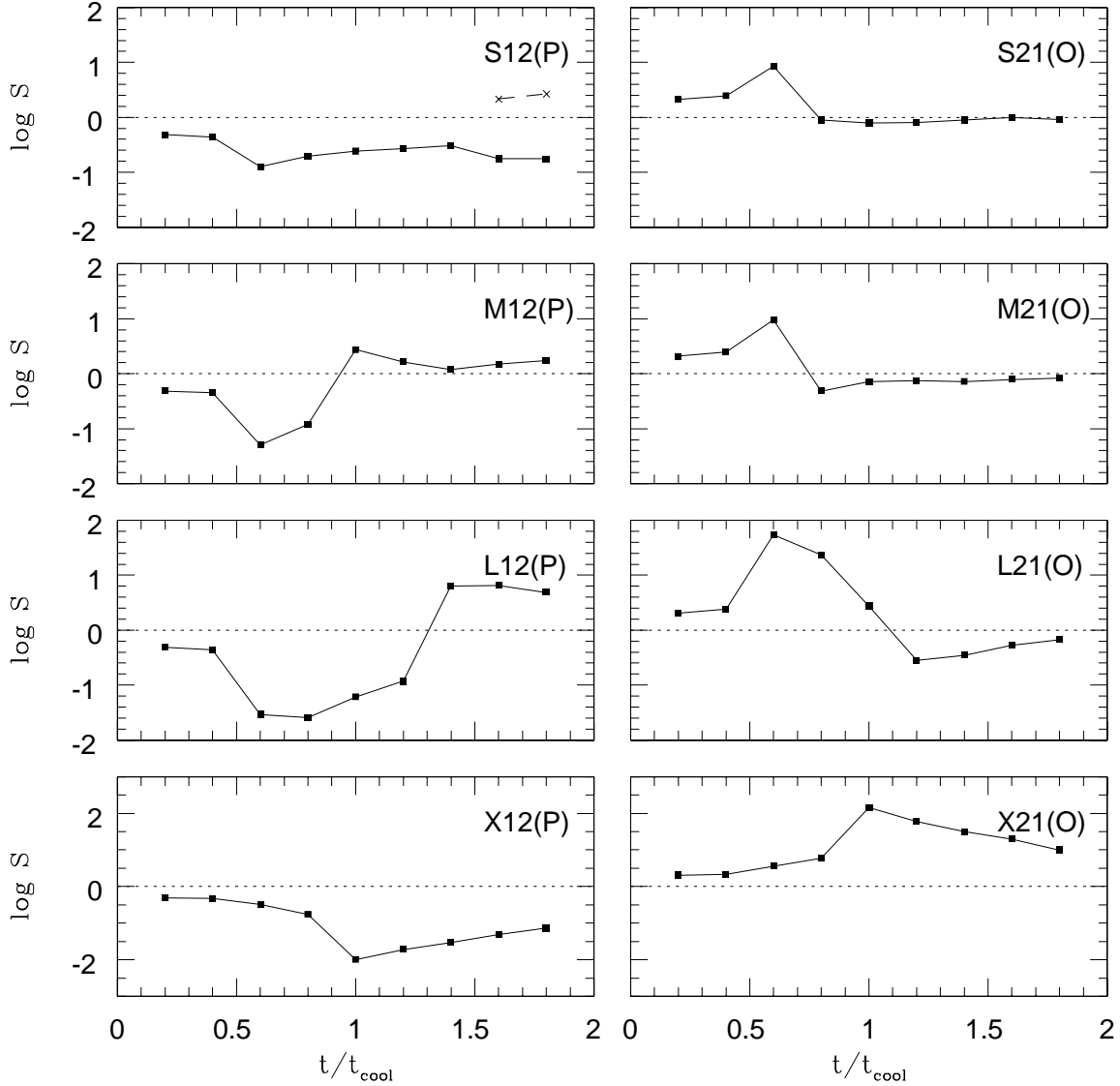}
\caption{
Time evolution of the shape parameter, $S$, for the prolate
(left panels) and oblate (right panels) clouds. The dotted horizontal
lines are applied to the spherical symmetry ($S=1$). In the S12 model,
the second shape parameter, $S_{100}$, is plotted for
$t\ge 1.6 t_{\rm cl,h}$ with a dashed line. 
}
\end{figure}
\begin{figure}
\plotone{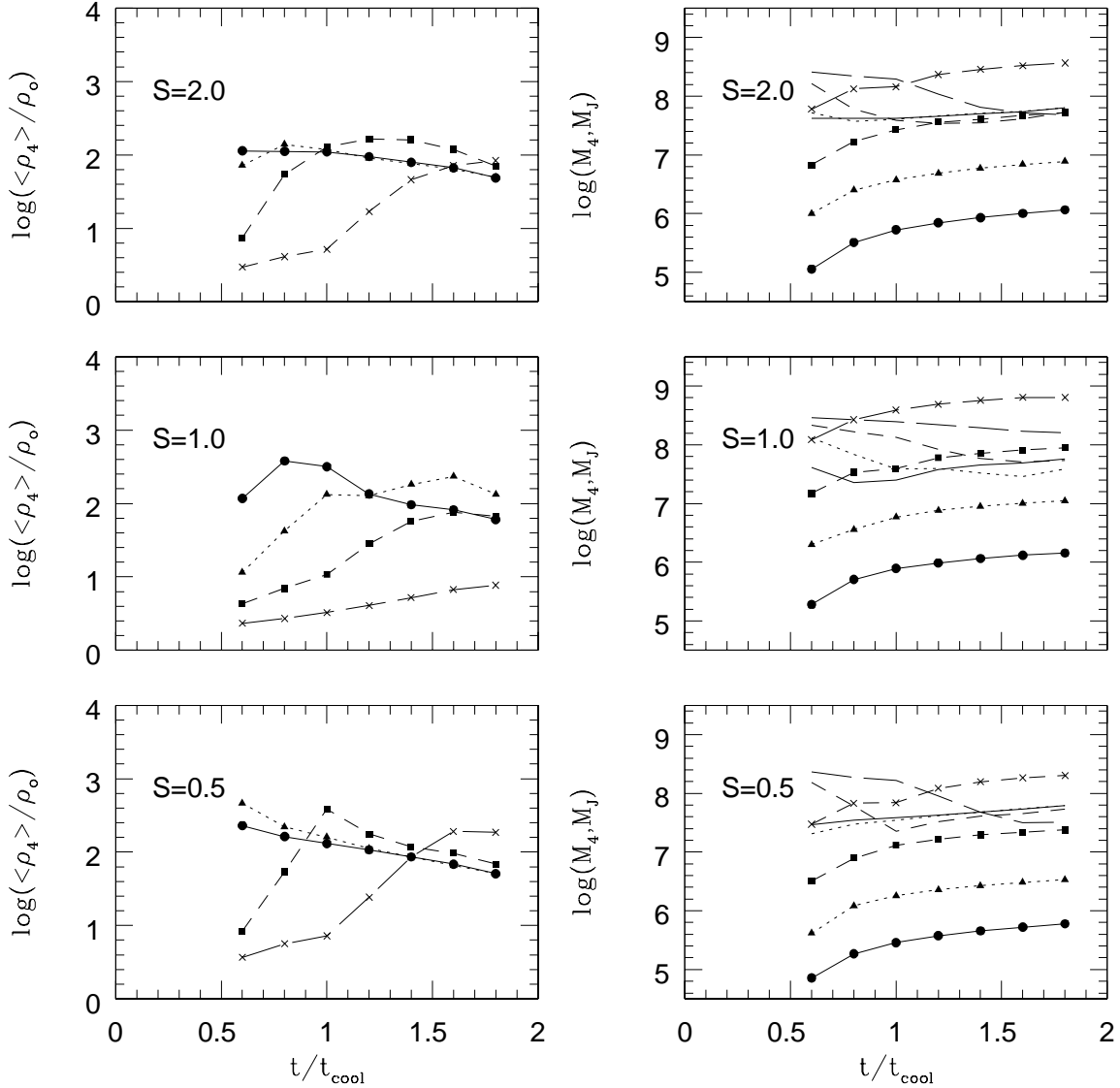}
\caption{
{\it Left panels}: Mean density of the gas with $T=10^4$K, $<\rho_4>$,
for prolate, spherical and oblate clouds (from bottom to top)
shown as a function of time.
The solid lines with circles are for the small cloud models,
the dotted lines with triangles for the medium size cloud models,
the dashed lines with squares for the large cloud models,
and the long-dashed lines with crosses for the very large cloud models. 
{\it Right panels}: Total mass of the gas with $T=10^4$K, $M_4$, shown
with the same line types and symbols as for $<\rho_4>$.
The spherical Jeans mass estimated with $T_c=10^4$K and $<\rho_4>$,
$M_J$, is also shown in right panels with the same line types but without
symbols.}
\end{figure}
\begin{figure}
\plotone{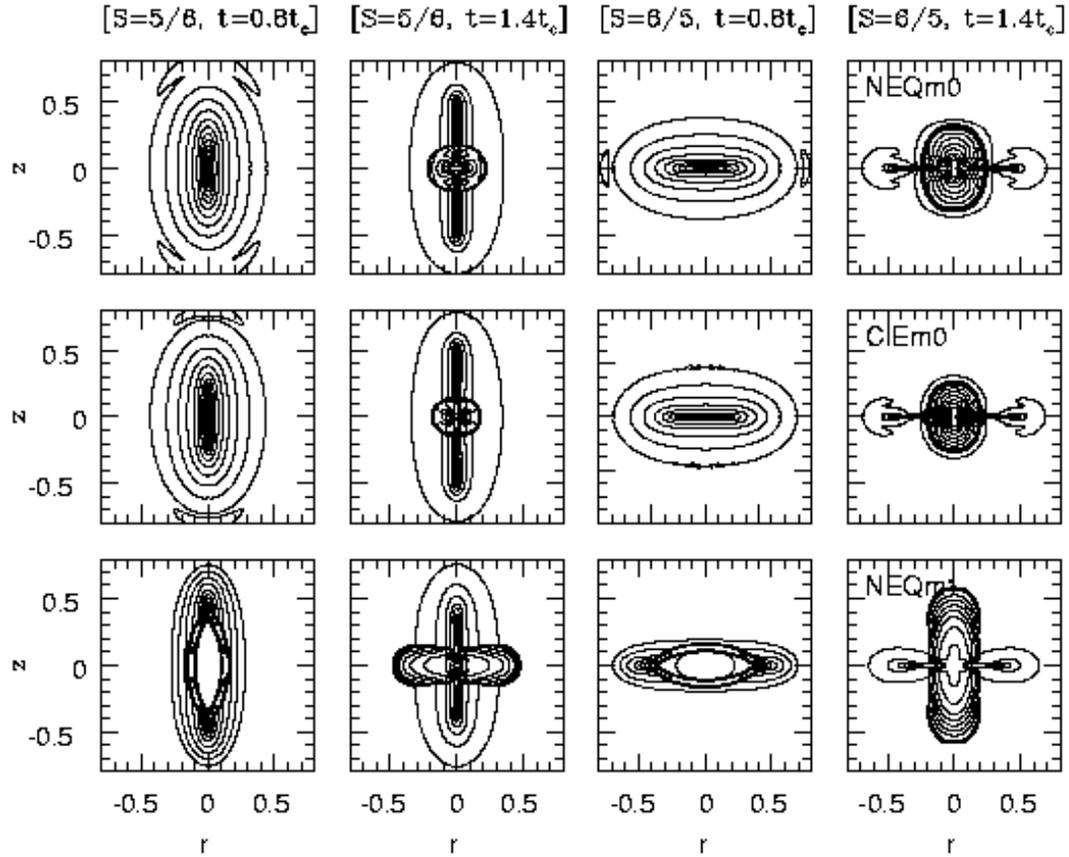}
\caption{
Density contour maps at $t = 0.8 t_{\rm cl,h}$ and $1.4 t_{\rm cl,h}$
for the large cloud models (L56 and L65) with three different cooling,
NEQm0, CIEm0, and NEQm1 (from top to bottom).}
\end{figure}

\end{document}